\RequirePackage{ifpdf}
\documentclass{PoS}
\usepackage[authoryear,square]{natbib}
\bibpunct{(}{)}{;}{a}{}{,}
\title{Unravelling the origin of large-scale magnetic fields in galaxy clusters and beyond through Faraday Rotation Measures with the SKA}

\ShortTitle{SKA: Magnetic fields in galaxy clusters and beyond}

\author{\speaker{Annalisa Bonafede$^{1}$}\thanks{On behalf of the SKA magnetism working group.}\\
        E-mail: \email{annalisa.bonafede@hs.uni-hamburg.de}}
\author{Franco Vazza$^{1}$, Marcus Br\"uggen$^{1}$, Takuya Akahori$^{2}$, Ettore Carretti$^{3}$, Sergio Colafrancesco$^{4}$, 
Luigina Feretti$^{5}$,
Chiara Ferrari$^{6}$, Gabriele Giovannini$^{7}$, Federica Govoni$^{8}$, Melanie Johnston-Hollitt$^{9}$, Matteo Murgia$^{8}$, 
Lawrence Rudnick$^{10}$, Anna Scaife$^{11}$,Valentina Vacca$^{12}$ \\
$^{1}$ Hamburger Sternwarte, Universit\"at Hamburg, Gojenbergsweg 112, 21029, Hamburg, Germany.
$^{2}$ Sydney Institute for Astronomy, School of Physics A28, University of Sydney, NSW 2006 Australia.
 $^{3}$     CSIRO Astronomy and Space Science, PO Box 76 Epping NSW 1710 Australia. 
 $^{4}$      University of the Witwatersrand, 1 Jan Smuts Avenue, Braamfontein 2000, Johannesburg 2000, South Africa.  
 $^{5}$      INAF Istituto di Radioastronomia, Via P. Gobetti, 101 40129 Bologna, Italy.        
  $^{6}$     Laboratoire Lagrange, UMR 7293, Observatoire de la c\^ote d'Azur, Boulevard de l'Observatoire, 06300 Nice, France.       
 $^{7}$        Dipartimento di Fisica e Astronomia, Viale Berti Pichat 6/2, Bologna University, Italy.         
 $^{8}$     INAF-Osservatorio Astronomico di Cagliari, Via della Scienza 5, 09047 Selargius (CA),  Italy.          
$^{9}$        Victoria University of Wellington, PO Box 600, Wellington, 6140, New Zeland.     
$^{10}$      Institute for  Astrophysics University of Minnesota, 116 Church St. SE Minneapolis, MN 55455, USA.     
 $^{11}$       University of Southampton, University Rd, Southampton SO17 1BJ, UK.       
 $^{12}$        Max Planck Institute for Astrophysics, Karl-Schwarzschild-Strasse 1, 85748 Garching bei M\"unchen, Germany.}

\abstract{We investigate the possibility for the SKA to detect and study the magnetic fields in galaxy clusters and in the less dense environments surrounding them using Faraday Rotation Measures. To this end, we produce 3-dimensional magnetic field  models for galaxy clusters of different masses and in different stages of their evolution, and derive mock rotation measure observations of background radiogalaxies. According to our results, already in phase I, we will be able to infer the magnetic field properties in galaxy clusters  as a function of the cluster mass, down to $10^{13}$ solar-masses.
Moreover, using cosmological simulations to model the gas density, we have computed the expected rotation measure through shock-fronts that occur in the intra-cluster medium during cluster mergers. The enhancement in the rotation measure due to the density jump will permit to  constraint the magnetic field strength and structure after the shock passage.
SKA observations of polarised sources located behind galaxy clusters will answer several questions about the magnetic field strength and structure in galaxy clusters, and its evolution with cosmic time.}

\FullConference{
Advancing Astrophysics with the Square Kilometre Array\\
June 8-13, 2014\\
Giardini Naxos, Italy}

\begin{document}

\section{Introduction}
\label{intro}
Magnetic fields are ubiquitous in the Universe but their origin is unknown.
On large scales, magnetic fields are the hardest to explain because the usually invoked dynamo mechanism does not have the time to amplify the field on the largest scales starting from a weak initial seed. In addition, the magnetic fields in galaxy clusters are poorly constrained from an observational point of view, and it is unclear whether they are formed from primordial  seeds - amplified during the process of structure formation -  or are formed from magnetic fields injected by AGN or galactic outflows.
The presence of magnetic field in galaxy clusters can be probed by diffuse radio sources associated with clusters (see contribution by Johnston-Hollitt et al., Govoni et al., Ferrari et al.),  and by Faraday rotation measures  (RM) of embedded and background polarised sources (see e.g. \citealt{2004IJMPD..13.1549G} for a review).  
 In the last decade, there have been  revolutionary improvements in  modelling the magnetic field in galaxy clusters, based partly on observations and partly on numerical simulations (e.g. \citealt{Bonafede11a}, \citealt{Bruggen13}, \citealt{Dolag08},  \citealt{2006A&A...460..425G}, \citealt{Vacca12}).
 Much of what is known about the magnetic field in galaxy clusters comes from sensitive polarised observations. The observed polarisation angle $\phi_{obs}$ of a synchrotron-emitting source observed in the background of a galaxy cluster is rotated with respect to the intrinsic polarisation angle by a quantity called Faraday Rotation Measure, defined as
$  RM=\int_{cluster} n_e(l) B_{//}(l) dl $,  where $n_e$ is the thermal gas density in the intra-cluster medium (ICM),  and $B_{//}$ is the magnetic field component along the line-of-sight.
 Recent studies have given us important clues to elucidate the evolution of the magnetic fields in galaxy clusters.
However, because of the limits of current instruments, further improvements will be hard to achieve.
A primary limiting factor  is the sensitivity of present facilities, that limits the feasibility of such studies to a few nearby  clusters where a sufficient number of background and embedded polarised sources can be detected. A secondary limiting factor is the small field-of-view of the instruments, which require multiple pointings and, hence, an enormous observing time to survey the entire cluster.
In this work, we analyse the capabilities of the SKA in studying the properties of magnetic fields in the ICM through a dense RM sampling of sources detected in the background of these objects. Using different approaches, we have produced mock RM observations
of galaxy clusters and of the regions around them. 
 In Sec. \ref{sec:method} we outline the method used to produce the RM mock observations, and in Sec. \ref{sec:clusters} we apply this method to clusters of different masses, magnetic field strengths and dynamical status. In Sec. \ref{sec:relic} we make use of cosmological simulations to explore the SKA capabilities in detecting the RM amplification expected in shocked regions. 
 We assume here that SKA1-SUR will be populated in band 1 (350-900 MHz) and  band 2 (650 -1670 MHz). 
This is motivated by the range of RM that we need to sample. Results do not change if SKA1-MID in used instead of SKA1-SUR. 
We assume that within each band the frequency channel width will be 1 MHz. 
 We discuss in Sec. \ref{sec:discussion} how our predictions change in view of the progressive development of the SKA.

\section{Mock RM observations}
\label{sec:method}
In order to predict the capabilities of the SKA in studying the magnetic fields on the large scales, we have produced mock RM
observations of single radiogalaxies detected through the ICM.
We have used an approach similar to the one described in \citet{Murgia04} and  \citet{Bonafede10}, that consists of modelling the gas distribution inside the galaxy cluster and the magnetic field in a separate 3-dimensional box. 
We define $\Lambda$ as the spatial scale in the real space, and the magnetic field power spectrum as  $P_B(\Lambda) \propto \Lambda^{n}$ within the range  $ \Lambda_{min} \leq \Lambda \leq \Lambda_{max}$.  We adopt here  $n=11/3$, corresponding to a Kolmogoorv scaling. The magnetic field is initially defined in the vector-potential domain. Through an appropriate choice of the vector potential components, we obtain  a magnetic field with Gaussian components, divergence-free, and isotropically oriented throughout the box.
The magnetic field is then normalised within the
core radius to a given value $B_0$, and is scaled according to $B \propto  B_0 \left( n_e /n_0 \right)^{\eta}$, 
$n_0$ being the cluster central gas density as defined in the $\beta$-model.
The free parameters of our magnetic field model are $B_0$, $\Lambda_{min}$, $\Lambda_{max}$, and $\eta$.  
For every realisation of the magnetic field and of the gas density profile (see Tab. \ref{tab:clusters}), we have integrated the RM equation to derive the Faraday RM  throughout the cluster. 
Extrapolating from the deep survey in the GOODS-N field, and following the sensitivity listed in the SKA science memo,  the SKA1-SUR is expected to detect 315 polarised sources per square degree  at 1.4 GHz\,  with 1 $\mu$Jy detection threshold and 1.6 arcsec resolution \citep{2014ApJ...785...45R}.
Assuming this number, we have computed the expected  RM for every polarised background source,
whose position is randomly extracted within a circle of given aperture radius, which is chosen depending on the mass of the simulated cluster
in order to trace the signal up to the distance where the cluster gives a substantial contribution.
This approach has the great advantage of using background objects only, and hence it minimises the possible local ICM contribution due to the interaction of cluster radio-galaxies with their environment. 
The magnetic field simulations are performed with the {\it MiRo'} code (see \citealt{Bonafede13} for further details).

\begin{table}

 \centering
   \caption{Parameters for the gas and the magnetic field in galaxy cluster simulations}
  \begin{tabular}{c c c c c c c  c c c}
  \hline         
Model  & Gas & M                  & $B_0$   & $\Lambda_{min}$ & $\Lambda_{max}$ & $\eta$ & $n_0$       &   $r_c$ & $\beta$\\
             &        & $M_{\odot}$ & $\mu$G & kpc                       &      kpc                   &              & cm$^{-3}$ & kpc     &                 \\
BM15\_3.9   &$\beta$Model &  10$^{15}$ & 3.9&  3 & 34  &  0.4 & 3.44$\times 10^{-3}$ & 290 & 0.75 \\
 BM15\_4.7 &$\beta$Model  &  10$^{15}$ & 4.7&  3 & 34    &  0.5& 3.44$\times 10^{-3}$ & 290 & 0.75\\
 BM15\_5.5  & $\beta$Model  &10$^{15}$ &  5.5 &  3 & 34  &  0.7& 3.44$\times 10^{-3}$ & 290& 0.75\\
 \hline
BM14\_1   &  $\beta$Model  & 10$^{14}$ & 1 &  3 & 17 &  0.5 & 3.44$\times 10^{-3}$ & 134 & 0.75\\
BM14\_3   &  $\beta$Model  & 10$^{14}$ & 3 &  3 & 17  &  0.5& 3.44$\times 10^{-3}$ & 134& 0.75\\
BM14\_5   &  $\beta$Model  & 10$^{14}$ & 5 &  3 & 17 &  0.5 & 3.44$\times 10^{-3}$ & 134& 0.75\\
BM13\_1     & $\beta$Model & 10$^{13}$ & 1 & 3 & 17  &  0.5& 3.44$\times 10^{-3}$ & 62 & 0.75\\
BM13\_3      & $\beta$Model & 10$^{13}$ &3 & 3 & 17  &  0.5& 3.44$\times 10^{-3}$ & 62 & 0.75\\
BM13\_5       & $\beta$Model & 10$^{13}$ & 5 & 3 & 17  &  0.5& 3.44$\times 10^{-3}$ & 62& 0.75\\
\hline
\hline
E1\_4     &  Enzo sim & 1.12$\times$10$^{15}$  & 4  &  34 & 115 &  0.5\\ 
E2\_4         &  Enzo sim  & 1.12$\times$10$^{15}$    & 4 & 34 & 115 &  0.5\\
E14\_4          & Enzo sim  &   10$^{15}$  & 4 & 34 & 115 &  0.5\\  

\hline
\end{tabular}
\label{tab:clusters}
\end{table}

\begin{figure*}
\vspace{85pt}
\begin{picture}(90,90)
\put(-20,-15){\includegraphics[width=8cm]{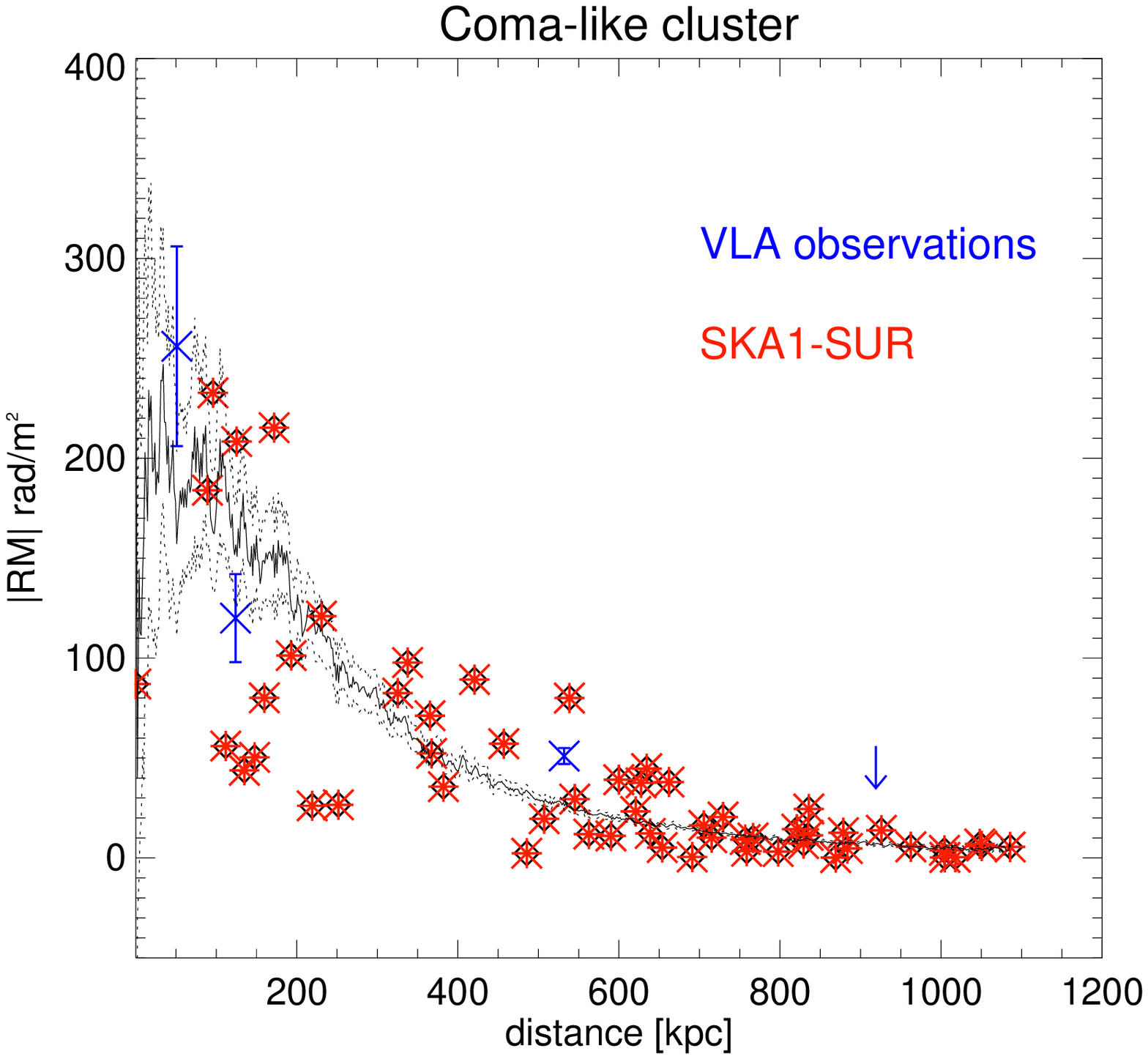}}
\put(200,-15){\includegraphics[width=8cm]{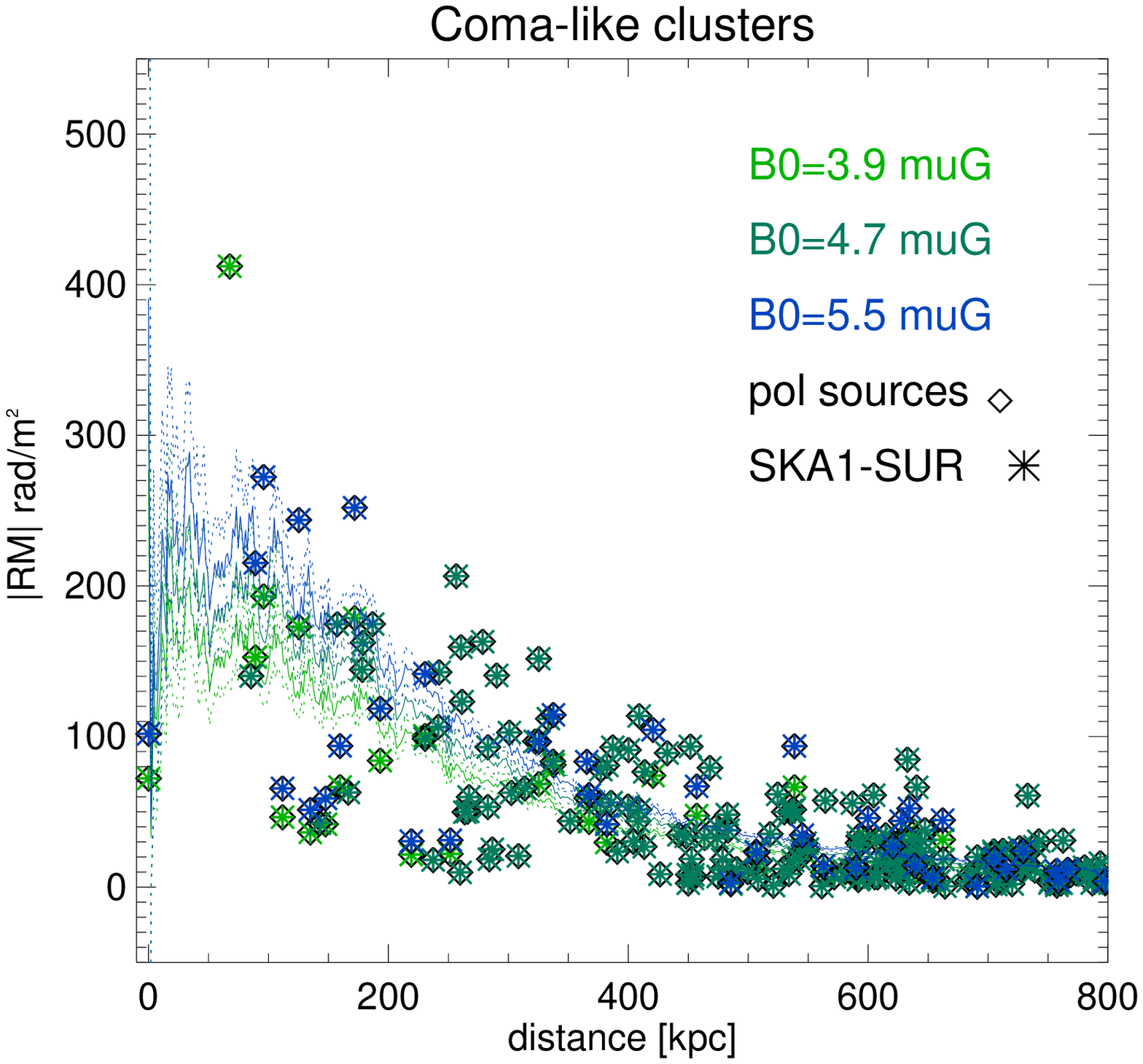}}
\end{picture}
\caption{SKA1-SUR predictions for the RM for Coma-like clusters. Left panel: Average profile of the RM (continuous line) and 5$\sigma$ dispersion (dotted lines) obtained for the BM15\_4.7 model.  Diamonds represent the sources that the SKA1-SUR will detect in the background of the cluster. Red asterisks mark the sources whose RM is in the range to be detected by the SKA1-SUR. Blue points refer to the observations presented in \citet{Bonafede10}.     Right Panel: lines and symbols are like in the left panel, shown in different colours for three different models as indicated in the inset.}
\label{fig:coma}
\end{figure*}

\begin{figure*}
\vspace{85pt}
\begin{picture}(90,90)
\put(-20,-15){\includegraphics[width=8cm]{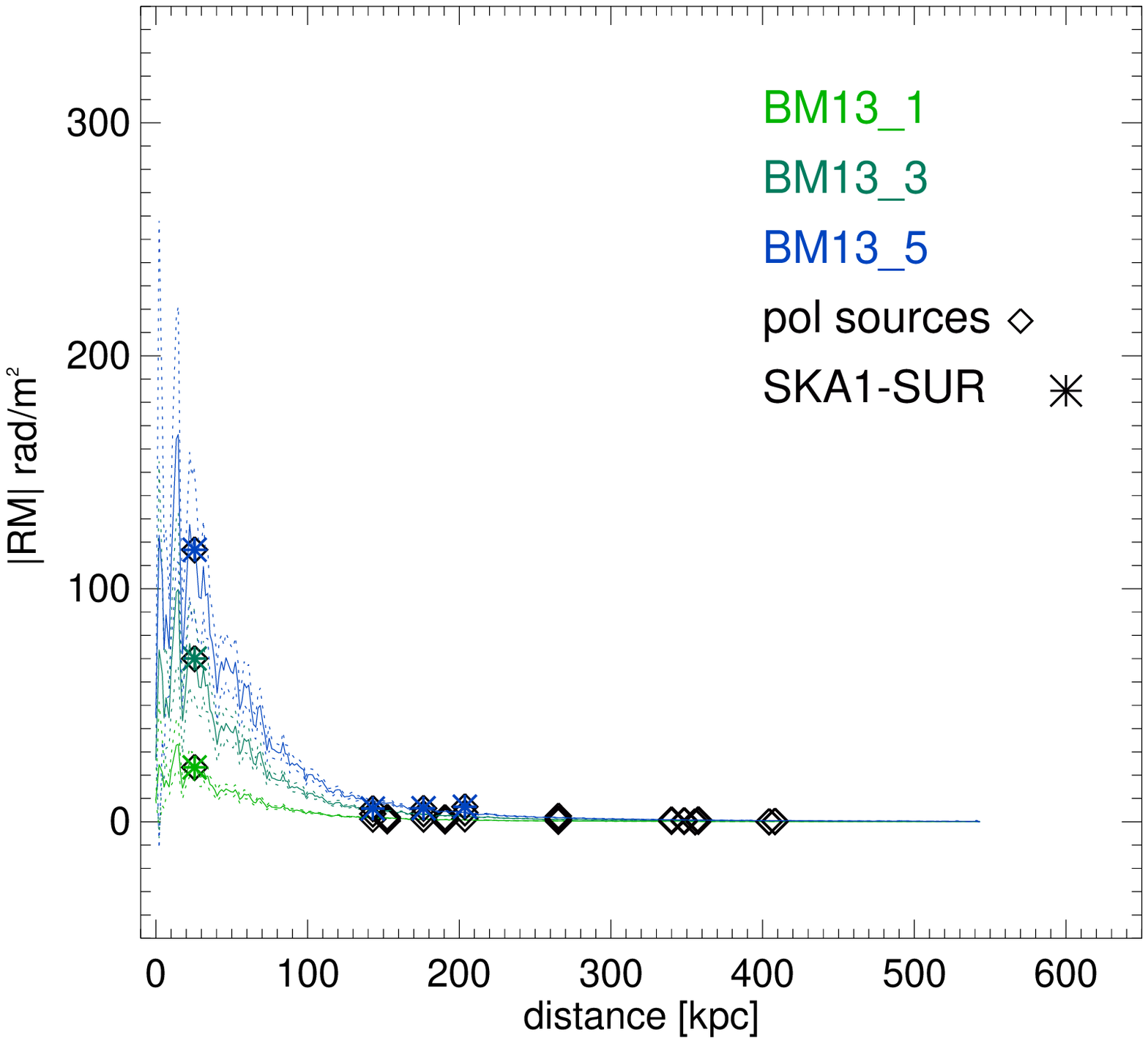}}
\put(200,-15){\includegraphics[width=8cm]{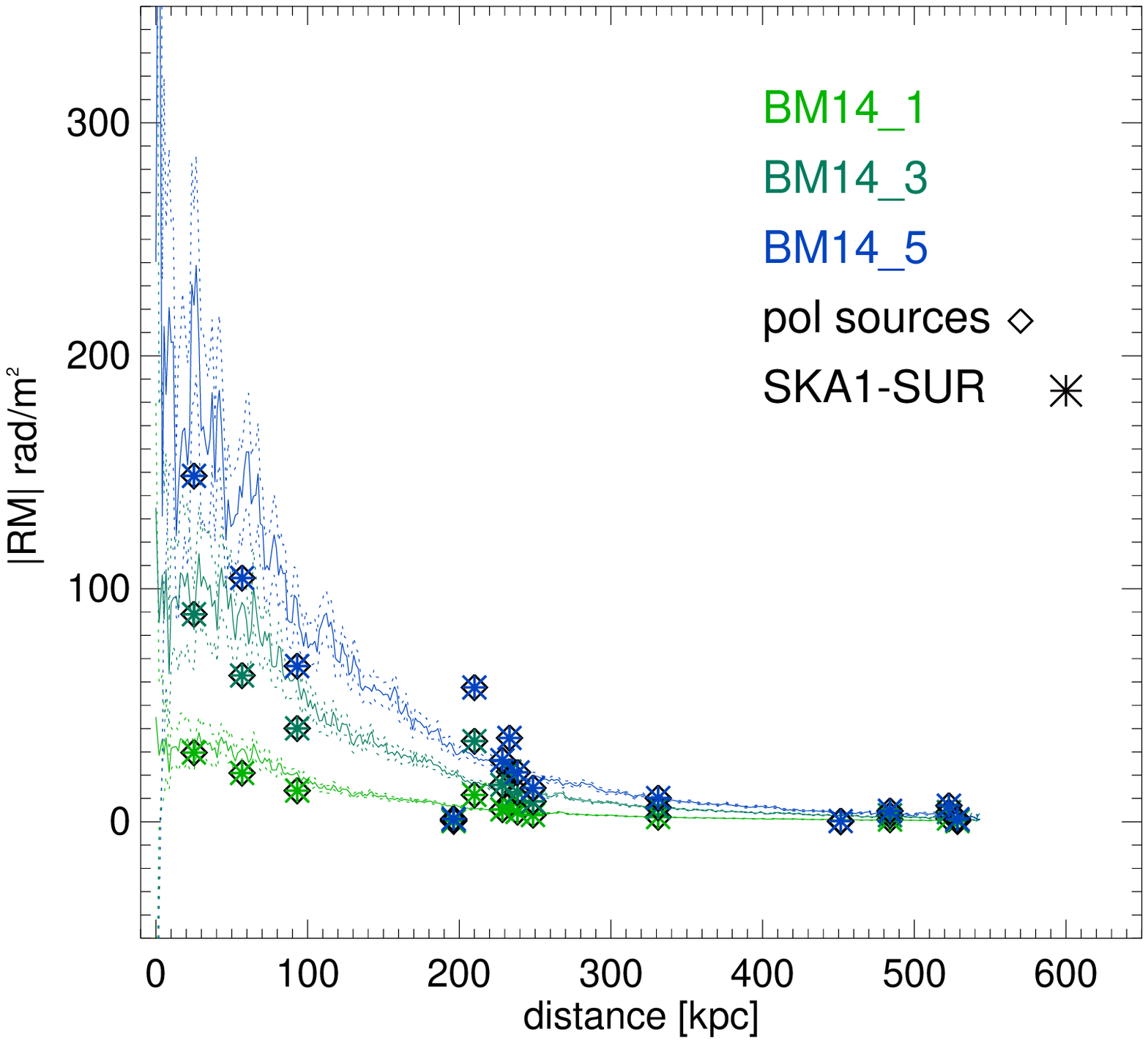}}
\end{picture}
\caption{SKA1-SUR predictions for clusters with $M=10^{13} M_{\odot}$ (left panel), and $M=10^{14} M_{\odot}$ (right panel). Lines and symbols are like in Figure 1.}
\label{fig:clusters}
\end{figure*}

\section{Magnetic field in galaxy clusters}
\label{sec:clusters}
In the following, we assume two distributions for the gas: (i) a $\beta$-model profile \cite{betamodel} and (ii) the gas distribution obtained  by a set of  cosmological simulations by \cite{va10kp}. 
\subsection{$\beta$-model profiles}
Our $\beta$-model clusters are meant to illustrate the big step forward allowed by the SKA in studying the magnetic field properties of galaxy clusters, compared to 
what has been achieved today with pointed observations of individual radiogalaxies.\\
The magnetic field in the Coma cluster is among the best constrained, and as such, it is a good starting point to investigate the
capabilities of the SKA compared to the present facilities. \\
We have considered a cluster gas density profile that  follows the $\beta$-model \citep{betamodel} derived for the Coma cluster by \citet{Briel01}   (models  BM15  in Table \ref{tab:clusters}). 
For $\Lambda_{min}$,  $\Lambda_{max}$,  we have adopted the parameters that give the best fit to the Coma cluster magnetic field profile, and we have simulated the 
different values of $B_0$ and $\eta$ that  give the best agreement with the observations at 68$\%$ confidence level \citet{Bonafede10}.
We have investigated the imprint that different values of $B_0$ and $\eta$  leave on the RM profile, to see if SKA1-SUR observations will be able to distinguish among these models. In Fig. \ref{fig:coma}  the RM profile obtained with SKA1-SUR observations  is shown for the model BM15\_4.7, a  "Coma-like" cluster. The results by \citet{Bonafede10}, obtained with VLA pointed observations, are overplotted.
SKA1-SUR  will allow one to detect the RM from $\sim$ 50 sources in the background of the "Coma-like" clusters. 
Thanks to the SKA1-SUR field-of-view and sensitivity, similar results for a single cluster will be achieved with a single pointing.\\
In Fig.  \ref{fig:coma}, we also show the RMs obtained by the SKA1-SUR for the different values of $B_0$ and $\eta$ that fit the VLA observations within the 68\% confidence level.
The resolution in RM of the SKA1-SUR will enable us, in principle, to distinguish among these models, although a more accurate evaluation of the errors needs to be done to definitively assess this point.\\
Another  aim of these $\beta$-model clusters  is  to show   how the SKA   will be able to recover the magnetic field profiles in relaxed systems depending 
on their mass and on the actual magnetic field strength at the cluster centre. So far, the magnetic field in galaxy clusters has been mainly analysed in massive and nearby  systems.
One of the reasons for this, is the number of sources that are needed in  the background or inside the cluster to perform a proper analysis.
To investigate how the SKA1-SUR will be able to constrain the magnetic field properties in clusters with different masses,
we have rescaled the $\beta$-model of the Coma cluster in a self-similar way,  for clusters with $10^{13}$ and 10$^{14}$ $M_{\odot}$, keeping the same values for 
$\Lambda_{min}$, and $\Lambda_{max}$, and changing the value of $B_0$, as specified in Table \ref{tab:clusters} (cluster models BM14 and BM13).
In Fig. \ref{fig:clusters} the predicted number of RM throughout clusters with different masses are shown, together with the RM profile derived from the
RM simulation. From these figures we can conclude that the sources relevant to establish the magnetic field properties are those within few hundred kpc
from the cluster centre. At larger distances ($\geq$ 300 kpc and 400 kpc for M13 and M14 models, respectively), the signal vanishes because of the lower gas density and magnetic field strength. Using only background sources
has the advantage of avoiding a  possible local RM enhancement. However, the information at the cluster centre, possibly provided by the radio-emitting
central source, will be likely missing. Nonetheless, the SKA will be the first instrument which provides a good sampling of the RM in the background of
smaller-mass clusters.

\begin{figure*}
\vspace{125pt}
\begin{picture}(130,130)
\put(15,5){\includegraphics[width=14cm]{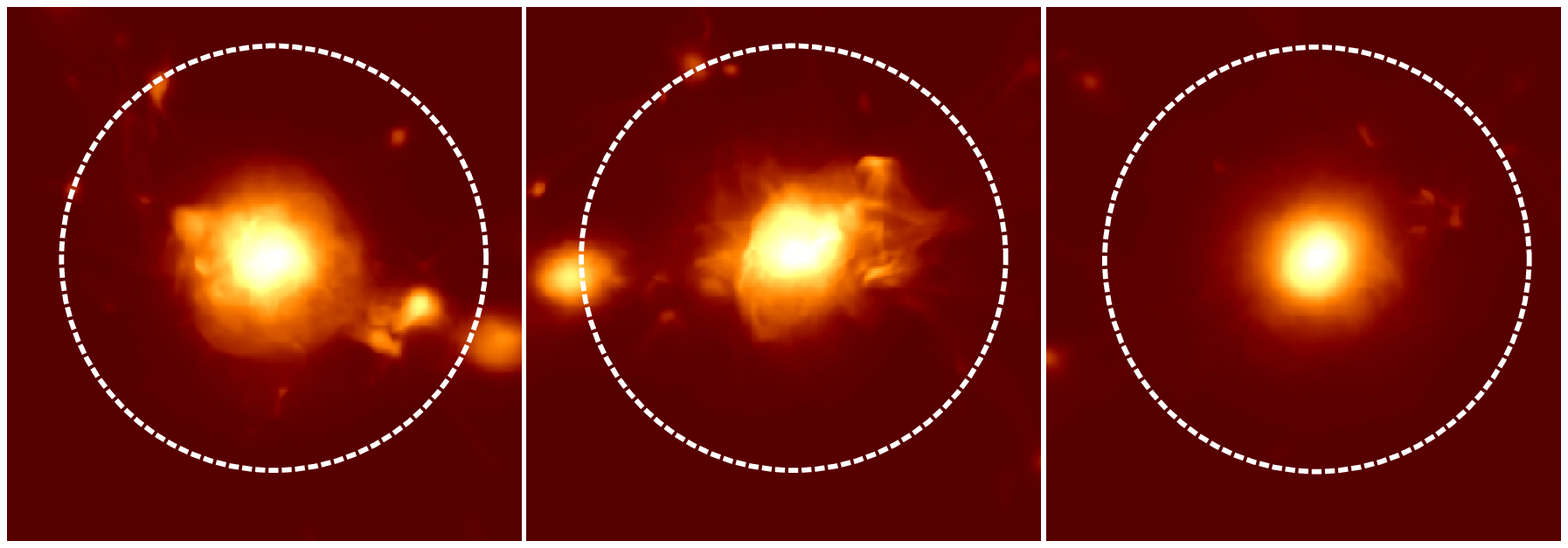}}
\put(0,150){\includegraphics[width=5cm]{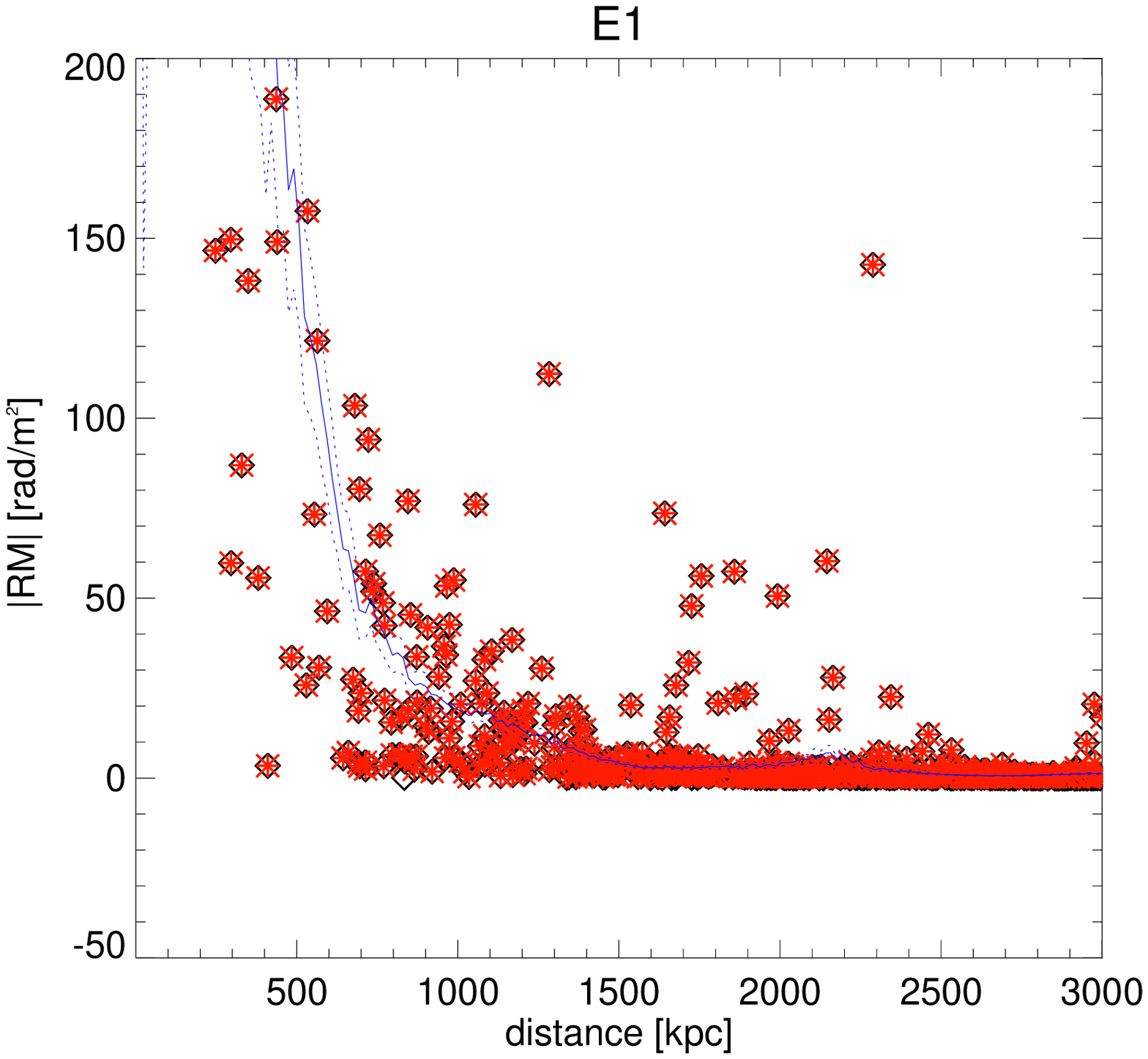}}
\put(140,150){\includegraphics[width=5cm]{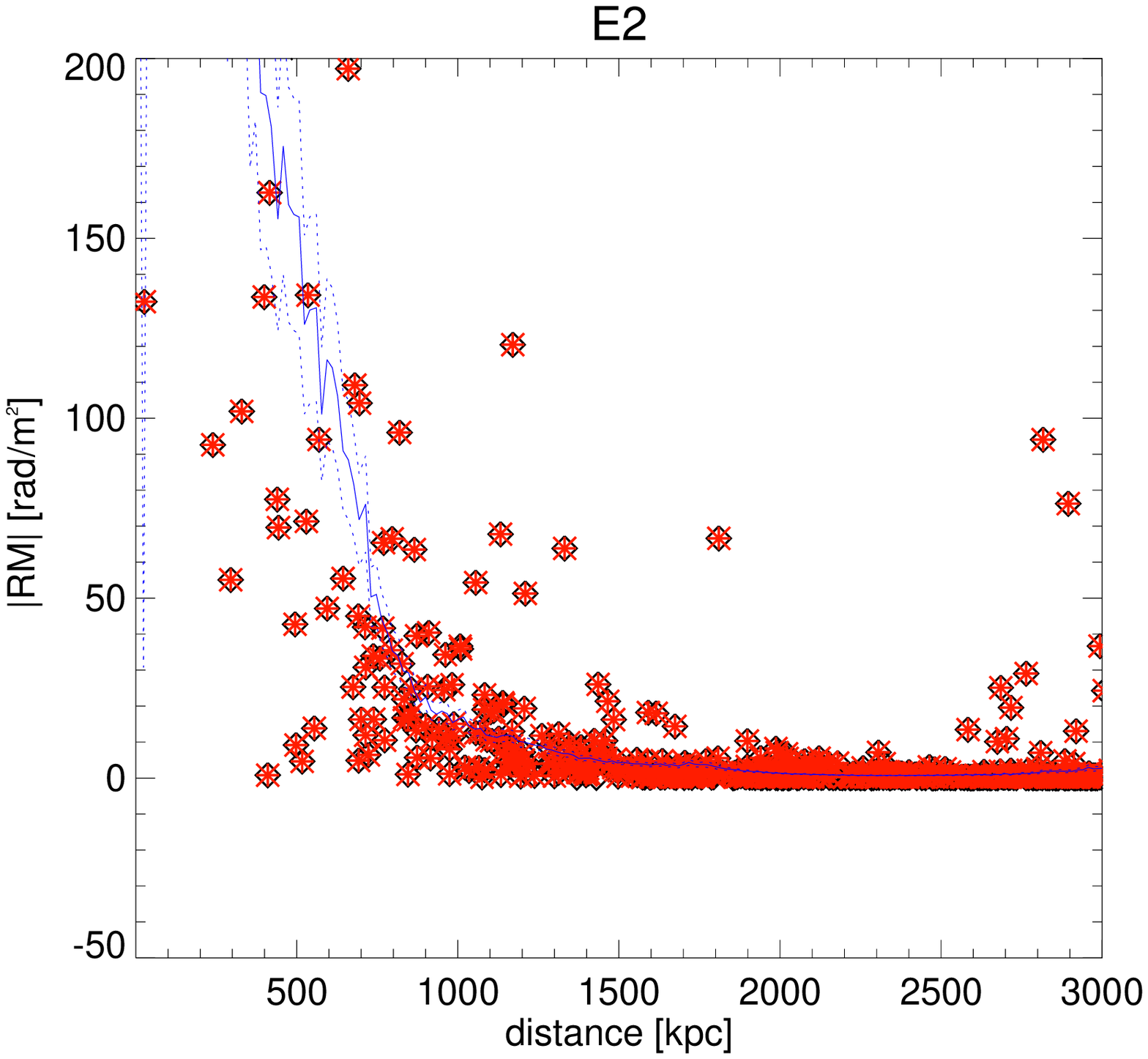}}
\put(275,150){\includegraphics[width=5cm]{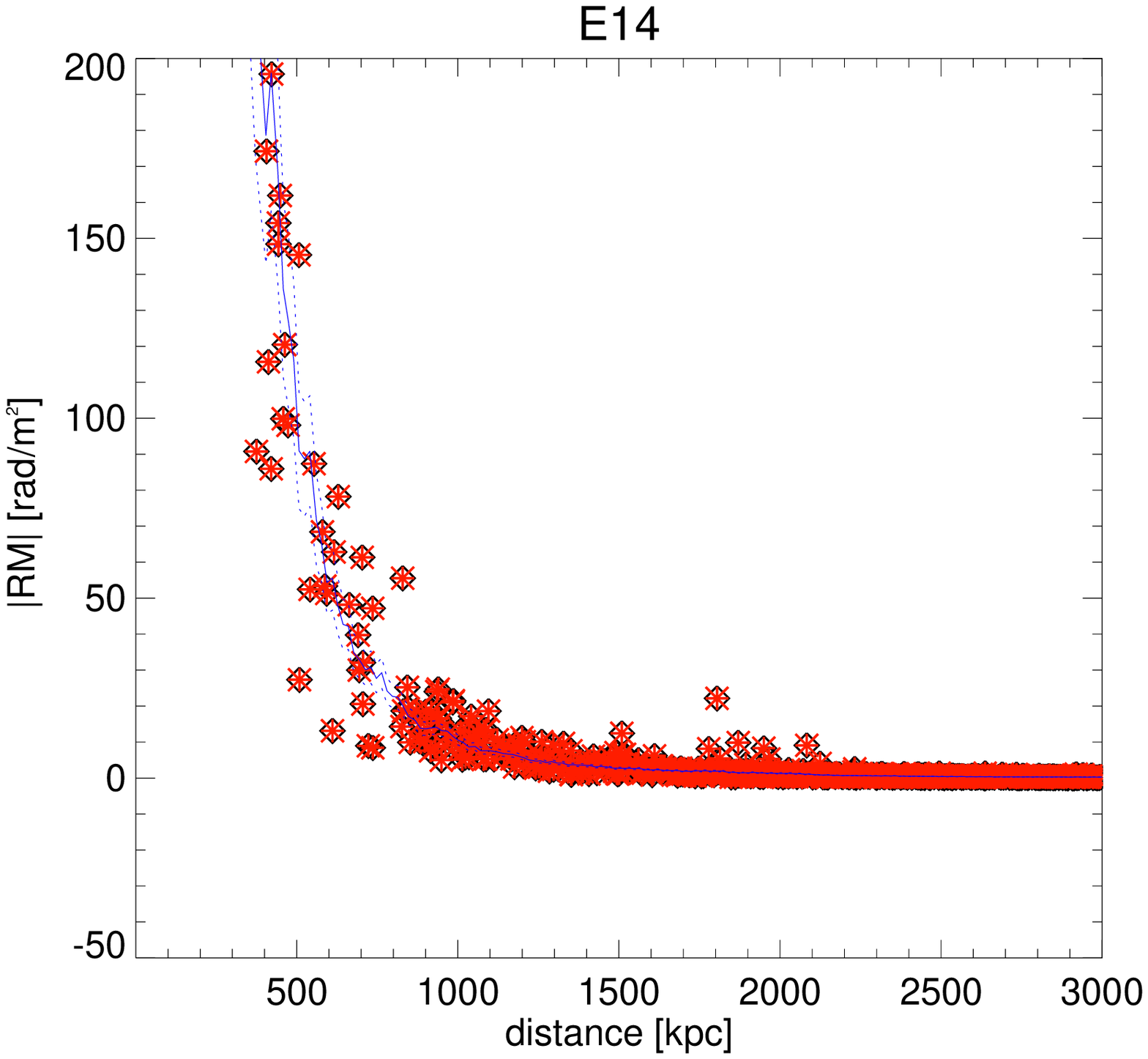}}
\end{picture}
\caption{ Top panel: RM profiles for the simulated clusters E1 (left), E2 (centre), and E14 (left). The y-axis goes up to 200 $rad/m^2$ in order to highlight the differences
among the different clusters. The blue continuous line represents the radial average of the RM profile, the dotted lines refer to the 5$\sigma$ dispersion. Red asterisks
mark the sources detected by the SKA1-SUR. 
Bottom panel: X-ray projected luminosity for the clusters shown above. The white dashed-circle is centred on the cluster X-ray peak
and has a radius of $\sim$ 1 virial radius. }
\label{fig:enzo}
\end{figure*}

\subsection{Cosmological simulations}
Although the $\beta$-model is often a good representation of the gas density profile in the ICM, high-resolution X-ray observations have shown
that the ICM is much more complex and difficult to characterise analytically (see e.g. \citealt{2009arXiv0906.4370B}, \citealt{Ebeling07}, \citealt{2013SSRv..177..119E}). Indeed, cosmological simulations of galaxy clusters
show that the matter is continuously accreting onto the cluster, generating turbulence, shock waves, and bulk motions in the ICM, which
leave a clear imprint in the gas density distribution. In order to investigate the capabilities of the SKA in  more complex and realistic environments,
we have repeated the procedure outlined above (Sec. \ref{sec:method})  starting from the gas density distribution obtained by cosmological simulations.
The sample of clusters by \citet{va10kp}  consist of 20 clusters, re-simulated with an Adaptive Mesh refinement
method optimised to have a high resolution in the outskirts of the clusters. The authors have classified the clusters  as  relaxed (RE), 
merging (ME), and post major-merger (MM).  Hence, we can  investigate the SKA1-SUR capabilities to recover the magnetic field properties in clusters that are in different 
phases of their evolution. In this work, we focus on  three different clusters, namely E1, E2 and E14, classified as  MM, ME, and RE, respectively.
The simulations have a resolution of 25 kpc h$^{-1}$. We have resampled the gas density distribution within the virial radius with a bigger grid, reaching a final resolution of 12 kpc h$^{-1}$.
Ideally, one would need a factor 10 better resolution to match the SKA1-SUR resolution at 1.4 GHz, but this cannot be obtained yet, because of computing
power limitations. Since the deviations from the $\beta$-model profile occur on scales larger than the resolution achieved here, we can still make useful forecasts for
the SKA in detecting magnetic fields in different dynamical environments. 
The magnetic field is attached to the gas as explained in Sec. \ref{sec:method}. We have fixed the parameters $B_0$, $\Lambda_{max}$,  $\Lambda_{min}$, and $\eta$ (see  Table \ref{tab:clusters}) and we have investigated the different trends of the RM in the different clusters. 
We have to note that the main limitation of our approach is the use of a magnetic field model, which does not evolve with the cluster. Hence, one can expect to detect only the main differences in the RM behaviour among the clusters.
In Fig. \ref{fig:enzo}, the X-ray luminosity of the clusters (from \citealt{va10kp}) together with the RM profiles are shown. From the RM mock observation, we have randomly extracted a number of sources that the SKA1-SUR will detect within one virial radius from the peak of the cluster X-ray emission. 
In all profiles, a decrease of the RM signal occurs at $\sim$ 1 Mpc (approximatively $r_{500}$ ) from the cluster centre. This decrease is more gradual for the clusters E1 and E2 with respect to the E14, which is relaxed. 
The main differences among E1, E2, and E14 arise at a distance of 500-3000 kpc, and are due to the presence of other clusters or sub-groups in the cases of E1 and E2, and to the almost empty environment which surrounds the relaxed cluster E14.
\begin{figure*}
\vspace{70pt}
\begin{picture}(80,80)
\put(10,0){\includegraphics[width=6cm]{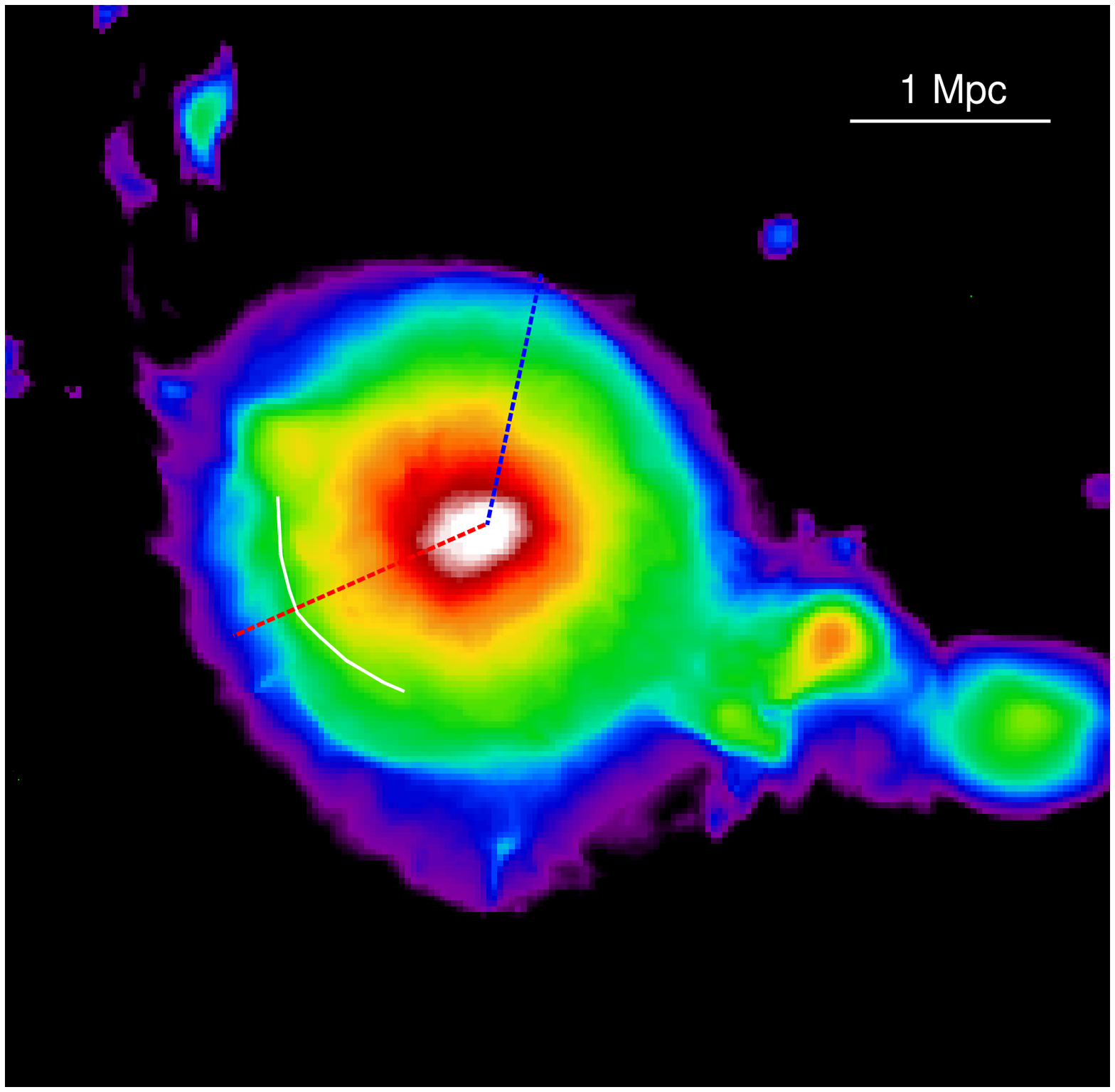}}
\put(210,-10){\includegraphics[width=7cm]{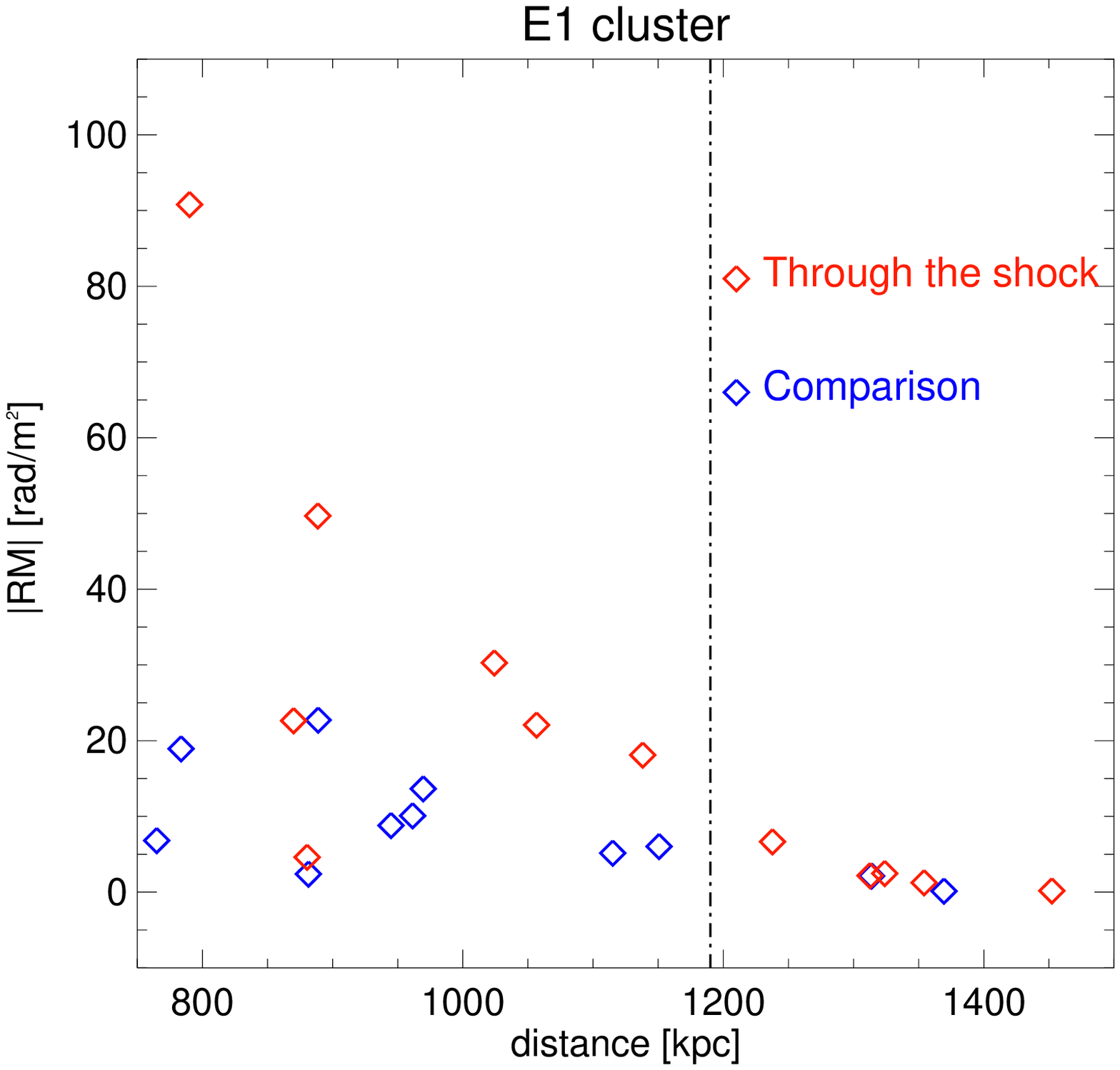}}
\end{picture}
\caption{Left panel: Colors display the X-ray luminosity of E1 in the [0.2-5] keV energy range \citep{va10kp}. The red and blue dashed lines indicate
the two directions considered to extract the RM sources, through a detected shock wave (red) and through a comparison region (blue). Right panel: RM of background sources extracted around the 2 directions shown in the left panel. The dotted-dashed line indicates the approximative position of the shock front. }
\label{fig:shock}
\end{figure*}

\section{Beyond galaxy clusters: magnetic fields and shock waves}
\label{sec:relic}
The use of cosmological simulations allows us to derive forecasts for the SKA in yet unexplored environments, such as 
the outskirts of galaxy clusters and intergalactic filaments (see Vazza et al, this contribution). The merger between massive galaxy clusters leads to the formation
of shock waves that travel through the ICM. Shock waves are supposed to be the origin of the so called radio-relic, a class of synchrotron sources
that are characterised by a low surface brightness, a steep radio spectrum, and a relatively strong polarisation (for details about radio relics we refer the reader to the reviews by \citealt{Bruggen11} and \citealt{feretti12}). Although the origin of radio relics is not well understood yet, if it understood that they are connected to shock waves. Hence, one expects to detect an enhancement of the RM signal due to the boost in the gas density and in the magnetic field by the shock. 
Such an enhancement has beed detected so far only for the relic in Abell 3667 \citep{2004rcfg.proc...51J},  while \citet{Bonafede13} have not found such a clear effect in the case of Coma.
The simulations by \citet{va10kp} have been generated with a refining method optimised to detect the velocity and temperature jumps
in the ICM. Here we focus of the cluster E1, where a shock front has been identified SE of the cluster centre.
Using the RM mock observation (model E1\_4), we have traced the RM profile in two regions: one across  the shock front, and the other
in a region where no shock has been detected. As done in the previous Sec., we have randomly extracted a number of sources that the SKA1-SUR will detect in polarisation.
In Fig. \ref{fig:shock} the two profiles are shown. Although the number of sources does not allow one to trace a profile and, possibly, detect a jump at the position of the shock front, a significant enhancement of the RM in the shocked region with respect to the comparison region is detected.
 Our magnetic field simulations would predict a magnetic field strength of $\sim$ 1 $\mu$G at the position of the shock front, which in fair agreement with the magnetic field strength derived through different methods for radio relics
\citep{feretti12}. A limitation of our approach is that the magnetic field is attached to the gas a posteriori, and is modelled with a simple recipe,  so that the magnetic field
compression and amplification expected at the shock position are not included in our mock observations.
Including these effects is tricky from a theoretical point of view, since the mechanism through which the different magnetic field components are amplified and ordered on large-scale is not known yet. As a result of this limitation, the enhancement that we obtain
has to be regarded as a lower limit to the effective RM enhancement that one would obtain once the compression and amplification of the shock are properly included. 
Remarkably, the values of the RM that we obtain match with the ones reported by \citet{2004rcfg.proc...51J}.
\section{Conclusions}
\label{sec:discussion}
We have explored the capabilities of the SKA in studying the properties of magnetic fields inside and around galaxy clusters. Our predictions indicate
that the SKA1-SUR will be able to recover tiny differences in the magnetic field properties of the ICM, which are far beyond the capabilities of the present instruments.
Thanks to the SKA 1-SUR sensitivity and resolution, we expect to detect hundreds of sources in the background of massive galaxy clusters, and to trace the RM enhancement due to the presence of merging shock waves.The results expected by the SKA in its ``early-science"  phase depend on the number counts of polarised sources detectable at a sensitivity which is $\sim$50\% of the  SKA1-SUR. The reduced number of polarised sources will not allow us yet to detect a sufficient number of RM sources behind small mass clusters ($M  \sim 10^{13} M_{\odot}$), and to detect a net enhancement through shocked regions on the ICM. Nonetheless, the results achieved here for massive galaxy clusters will be reachable already in the ``early-science" phase, though with lower precision.
The SKA2 will have a sensitivity ten times better than the SKA1, and a field of view corresponding to 20 times the one of the SKA1. Although a precise forecast for the number counts of polarised sources is not available yet, even in the most pessimistic case, that the number counts of polarised sources will be the same, the
SKA2 will allow to obtain the results presented here in a shorter observing time.  

\section*{Acknowledgements}
A.B. F.V and M.B acknowledge support from the grant FOR1254 from the Deutsche Forschungsgemeinschaft. 
F.V. acknowledges computational resources under the CINECA-INAF 2008-2010 agreement, and
the  usage of computational resources on the JUROPA cluster at the at the Juelich Supercomputing Centre (project no. 5963).
MJ-H acknowledges support from the Marsden Fund.
\bibliographystyle{apj}
\bibliography{master}

\end{document}